\documentclass[12pt]{article}
\usepackage{amssymb}
\usepackage{amsmath}
\usepackage{epsfig}
\usepackage{float}
\newcommand{\be}{\begin{equation}}
\newcommand{\ee}{\end{equation}}
\newcommand{\bea}{\begin{eqnarray}}
\newcommand{\eea}{\end{eqnarray}}
\begin{document}

\begin{center}
{\bf NEUTRINOLESS DOUBLE $\beta$-DECAY: STATUS AND FUTURE }
\footnote{
A report at the International conference ``Non-Accelerator New Physics'' NANP05
Dubna, Russia, 
June 20-25, 2005.}
\end{center}

\begin{center}
S. M. Bilenky
\end{center}

\begin{center}
{\em  Joint Institute
for Nuclear Research, Dubna, R-141980, Russia, and\\
SISSA, I-34014 Trieste, Italy.}
\end{center}
\begin{abstract}
A brief summary of the status of neutrino masses, mixing and
oscillations is presented. Neutrinoless double $\beta$-decay is
considered. Predictions for the effective Majorana mass are
reviewed. A possible test of the calculations of
nuclear matrix elements  of the $0\nu\beta\beta$-decay is proposed.
\end{abstract}

\section{Introduction}
Observation of neutrino oscillations in the Super-Kamiokande \cite{SK} , SNO \cite{SNO}
KamLAND \cite{Kamland}, K2K \cite{K2K} and other neutrino experiments 
\cite{Cl,Gallex,Sage,SKsol,Soudan,Macro} is one of the most important
recent discovery in particle physics. There are no natural
explanations of the smallness of neutrino masses in the framework of
the Standard Model. A new,  beyond the SM mechanism of neutrino mass
generation is required. Several such mechanisms were proposed (see \cite{Theory}). In
order to ensure a progress in the understanding of the origin of
small neutrino masses and peculiar neutrino mixing new experimental
data are definitely necessary.

One of the most important problem which must be addressed  by the
future experiments is the problem of the nature of massive
neutrinos: are they Dirac or Majorana particles? Investigation of
the neutrinoless double $\beta$-decay ($0\nu\beta\beta$-decay) 
is the most effective method which could allow
to resolve this dilemma.

We will review here the status of the $0\nu\beta\beta$-decay.
Calculation of the nuclear matrix elements of the process is a
complicated theoretical problem. We will discuss here a possible
method which could allow to check the calculations.

We will  present first a brief summary of the neutrino oscillations
(theory and experimental data).

The theory of the neutrino oscillations  (see \cite{BPet,BGG} ) is based on the assumption that
field $\nu_{l L}(x)$ in the standard charged and neutral
currents
\be
 j^{CC}_{\alpha}(x)= 2 \sum_{l=e, \mu, \tau}\bar\nu_{l L}(x)
\,\gamma_{\alpha}\,l_{L}(x);~ j^{NC}_{\alpha}(x) = \sum_{l=e, \mu,
\tau}\bar \nu_{l L}(x)\,\gamma_{\alpha}\,\nu_{l L}(x)\label{1} 
\ee
are "mixed" fields 
\footnote{ We are assuming  here that the number of massive neutrinos is equal to the number of flavor neutrinos (three). All existing data (with the exception of the data of the LSND experiment 
\cite{LSND}) are in a perfect agreement with this assumption. The data of the LSND experiment will be checked by the running MiniBooNE experiment \cite{Miniboone}.}
\be 
\nu_{l L}(x) = \sum_{i=1}^{3}U_{{l}i} \, \nu_{iL}(x)
;~~l=e,\mu,\tau.
\label{2} 
\ee
Here $U$ is PMNS \cite{Pont,MNS} mixing matrix, $\nu_{i}$ is the field of neutrino
with mass $m_{i}$.

From Eqs. (\ref{1}) and (\ref{2}) follows  that flavor lepton numbers
$L_{e}$, $L_{\mu}$ and $L_{\tau}$ are not conserved. If total lepton
number $L=L_{e}+L_{\mu}+L_{\tau}$ is conserved, $\nu_{i}(x)$ is 
Dirac field of neutrinos $\nu_{i}$ ($L=1$) and antineutinos
$\bar\nu_{i}$ ($L=-1$).

If total lepton number $L$ is not conserved, $\nu_{i}(x)$ is the field of
Majorana neutrinos. The Majorana field $\nu_{i}(x)$ satisfies the condition
\be 
\nu^{c}_{i}(x)=C\,\bar\nu^{T}_{i}(x) =\nu_{i}(x), \label{3} 
\ee
where $C$ is the matrix of the charge conjugation.

The state of  the flavor neutrino
$\nu_{l}$ ($l=e,\mu,\tau$ ),  produced in a CC weak process
together with the lepton $l^{+}$, is described by  "mixed'' vector of state
\be
|\nu_{l}\rangle =\sum^{3}_{i=1}U^{*}_{li}\,|\nu_{i}\rangle. \label{4}
\ee
Here $|\nu_{i}\rangle$ is the state of neutrino with mass $m_{i}$,
momentum $\vec{p}$ and energy

$$E_{i}\simeq p + \frac{m^{2}_{i}}{2p}$$

Let us assume that  at $t=0$  flavor neutrinos $\nu_{l}$  with momentum $\vec{p}$ are produced. 
At
the time $t$ for the neutrino state we have
\be|
\nu_{l}\rangle_{t}
=\sum_{i}e^{-iE_{i}\,t}\,U^{*}_{li}\,|~\nu_{i}\rangle=
\sum_{l'}|~\nu_{l'}\rangle
\sum_{i}U_{l'i}\,e^{-iE_{i}\,t}\,U^{*}_{li}
 \label{5} 
\ee

Probability of the transition $\nu_{l} \to \nu_{l'}$ is given by
\be
{\mathrm P}(\nu_{l} \to \nu_{l'}) =|~\delta_{l'l}+ \sum_{i\geq
2}U_{l' i} \,~( e^{- i\,\Delta m^2_{1i} \frac {L}{2E}}-1) \,~U_{l
i}^*~|^2.\label{6}
\ee
Here $\Delta m^2_{ik}=m^2_{k}- m^2_{i}$ and $L\simeq t$ is the
distance between neutrino production and detection points. In the general case the
probability ${\mathrm P}(\nu_{l} \to \nu_{l'})$ depends on six
parameters: two mass-squared differences $\Delta m^2_{12}$ and
$\Delta m^2_{23}$, three mixing angles $\theta_{12}$, $\theta_{23}$,
$\theta_{13}$ and $CP$ phase $\delta$.

From the data of the neutrino oscillation experiments  it was found that
\begin{enumerate}
\item
$\Delta m^2_{12} \ll \Delta m^2_{23}.$

\item The angle $\theta_{13}$ is small.
\end{enumerate}
From analysis of the data of the reactor CHOOZ experiment 
\cite{CHOOZ} it was
obtained that
\be
\sin^{2}\theta_{13}\lesssim 5 \cdot 10^{-2}\label{7}
\ee
If we neglect in transition probabilities  small terms proportional to 
$\frac{\Delta m^2_{12} }{\Delta m^2_{23} }$ and
$\sin^{2}\theta_{13}$, then in the region of
$\frac{L}{E}$ sensitive to $\Delta m^2_{23}$ (atmospheric and
accelerator long baseline  experiments) dominant transitions are
(see \cite{BGG})
$$\nu_{\mu} \rightleftharpoons \nu_{\tau}~~
(\bar\nu_{\mu} \rightleftharpoons \bar\nu_{\tau})$$

Probability of $\nu_{\mu}$ ($\bar\nu_{\mu}$) to survive is given by
\be
{\mathrm P}(\nu_\mu \to \nu_\mu) = {\mathrm P}(\bar\nu_\mu \to
\bar\nu_\mu)= 1 - \frac {1} {2}\,\sin^{2}2\theta_{23}\, (1-\cos
\Delta m_{23}^{2}\, \frac {L} {2E}).\label{8}
\ee
The data of the atmospheric Super-Kamiokande experiment are
perfectly described by Eq. (\ref{8}). For the parameters $\Delta
m^{2}_{23}$ and $\sin^{2}2 \theta_{23}$ the following 90 \%~CL
ranges were obtained \cite{SK}:
\be
1.5\cdot 10^{-3}\leq \Delta m^{2}_{23} \leq 3.4\cdot
10^{-3}~\rm{eV}^{2};~~ \sin^{2}2 \theta_{23}> 0.92. \label{9}
\ee
In the regions  sensitive to $\Delta m^{2}_{12}$ (solar, KamLAND
experiments) effects of "large" $\Delta m^{2}_{23}$ is averaged. In
the leading approximation vacuum probability of $\bar\nu_{e}$ to
survive is given by
\be
{\mathrm P}(\bar \nu_e \to \bar\nu_e)
=1-\frac{1}{2}~\sin^{2}2\,\theta_{12}~ (1 - \cos \Delta m_{12}^{2}
\,\frac {L}{2E}).\label{10}
\ee
The probability of $\nu_{e}$ to survive in  matter is given by
\be
{\mathrm P}(\nu_e \to \nu_e)= {\mathrm P}^{\rm{mat}}_{\nu_e \to
\nu_e}(\Delta m_{12}^{2},~ \sin^{2}\theta_{12},~
\rho_{e}),\label{11}
\ee
where ${\mathrm P}^{\rm{mat}}_{\nu_e \to \nu_e}$ is two-neutrino
probability of $\nu_{e}$ to survive in  matter ($\rho_{e}$ is the
electron density). From global analysis of solar and KamLAND data
for the parameters $\Delta m_{12}^{2}$ and $\sin^{2}\theta_{12}$ it
was found \cite{SNO}
\be  
\Delta m^{2}_{12} =( 8.0^{+0.6}_{-0.4})~10^{-5}~\rm{eV}^{2};~~
\tan^{2} \theta_{12}= 0.45^{+0.09}_{-0.07}\label{12} 
\ee
Information on the lightest neutrino mass $m_{0}$ can be inferred
from the measurement of the high-energy part of the $\beta$-
spectrum of tritium. From the data of Mainz \cite{mainz} and Troitsk \cite{troitsk} experiments
the following upper bound was obtained
\be 
m_{0} \leq 2.3 ~\rm{eV}\label{13}
\ee
From several analysis of cosmological data, in which results of different measurements 
were used,  for the sum of the
neutrino masses upper bounds in the range
\be 
\sum_{i} m_{i} \leq(0.4-1.7)~ \rm{eV}\label{14}
\ee
were deduced \cite{Tegmark}.

Investigation of neutrino oscillations can not reveal the nature of
neutrinos with definite masses $\nu_{i}$ \cite{BHP}. In fact, Majorana and
Dirac mixing matrices are connected by the relation
\be 
U^{M}=U^{D}\,S(\beta),\label{15}
\ee
where $ S_{ik}(\beta)=e^{i\,\beta_{i}}\,\delta_{ik}$ is phase matrix
($\beta_{3}=0$). From (\ref{6})
and (\ref{15}) it is obvious that phase matrix $S(\beta)$ drops out
from the expression for the transition probability. We have
\be
{\mathrm P}^{M}(\nu_{l} \to \nu_{l'})=
{\mathrm P}^{D}(\nu_{l} \to \nu_{l'}).\label{16}
\ee

\section{Neutrinoless double $\beta$-decay}

The search for neutrinoless double $\beta$-decay
\be 
(A,Z) \to (A,Z+2)+ e^{-}+ e^{-} \label{17} 
\ee
of $^{76} \rm{Ge} $, $^{130} \rm{Te} $, $^{136} \rm{Xe} $, $^{100}
\rm{Mo}$ and other even-even nuclei is the most sensitive and direct
way of the investigation of the nature of neutrinos with definite
masses $\nu_{i}$ (see reviews \cite{bbreviews1,bbreviews2}).  The total lepton number in $0\nu\beta\beta$ -decay
is violated and the process is allowed only if  $\nu_{i}$ are Majorana
particles \cite{Valle}.

We will discuss now the process (\ref{17}). The  effective
Hamiltonian of the process is given by 
\be 
\mathcal{H}_{I}^{\mathrm{CC}}= \frac{G_{F}}{\sqrt{2}}\,2 \bar
e_{L}\gamma _{\alpha}\nu_{eL}\,~ j^{\alpha} + \mathrm{h.c.}\,.
\label{18} 
\ee

Here $j^{\alpha}$ is the hadronic charged current, $G_{F}$ is the
Fermi constant and
\be \nu_{eL} = \sum_{i} U_{ei} \nu_{iL}~, \label{19} \ee
where $\nu_{i}$ are Majorana fields.

The neutrinoless double $\beta$-decay  is the second order in
$G_{F}$ process with virtual neutrinos. The neutrino propagator is
given by the expression
\be 
<0|T(\nu_{eL}(x_{1})~\nu^{T}_{eL}(x_{2}))|0> =
m_{ee}~\frac{i}{(2\,\pi)^{4}}\int {e^{-ip(x_{1}-x_{2})}\,
\frac{1}{p^{2}-m^{2}_{i} } \frac{1-\gamma_{5}}{2} \,d^{4}p\,C.} \label{20}
\ee
Here
\be 
m_{ee} = \sum_{i}U^{2}_{ei}\,m_{i}. \label{21} 
\ee
is {\em effective Majorana mass}. For small neutrino masses 
$m^{2}_{i}\ll p^{2}$ and we can safely neglect $m^{2}_{i}$ in the
denominator of the neutrino propagator.
Therefore, the matrix element of the
$0\nu\beta\beta$ -decay is factorized in the form of a product of
the effective Majorana mass, which is determined by neutrino masses $m_{i}$
and $U^{2}_{ei}$,  and nuclear
matrix element, which is determined only by nuclear properties and
does not depend on neutrino masses and mixing.

The half-life of $0\nu\beta\beta$ decay $T^{0\,\nu}_{1/2}(A,Z)$ is
given by the expression 
\be
\frac{1}{T^{0\,\nu}_{1/2}(A,Z)}=
|m_{ee}|^{2}~|M^{0\,\nu}(A,Z)|^{2}~G^{0\,\nu}(E_{0},Z),\label{22}
\ee
where  $M^{0\,\nu}(A,Z) $ is nuclear matrix element (NME) and
$G^{0\,\nu}(E_{0},Z)$ is known phase-space factor.

The results of several experiments on the search for
$0\nu\beta\beta$ -decay are available at present. No  commonly accepted evidence in
favor of the decay was obtained so far.\footnote{ The recent claim \cite{Klap}
of evidence of the  $0\nu\beta\beta$ -decay of $^{76} \rm{Ge}$  will
be checked by the GERDA experiment \cite{gerda}.}
The most stringent lower bounds for the half-time of the
${0\nu\beta\beta}$- decay were obtained in the Heidelberg-Moscow \cite{HM} and
CUORICINO \cite{Cuoricino} experiments.

In the Heidelberg-Moscow experiment  the following lower bound for
the half-life of $^{76} \rm{Ge}$ was found
\be 
T^{0\nu}_{1/2}\geq 1.9 \cdot 10^{25}\, \rm{years}\label{23}
\ee

In the CUORICINO experiment for half-life of
$^{130} \rm{Te}$ the following bound was obtained
\be 
T^{0\nu}_{1/2}\geq 1.8 \cdot 10^{23}\, \rm{years}\label{24}
\ee
Taking into account different calculations of the nuclear matrix
elements, for the effective Majorana mass from these results the
following upper bounds were deduced
\be 
|m_{ee}| \leq (0.3-1.2)~\rm{eV}~(\rm{H-M});~|m_{ee}| \leq
(0.2-1.1)\,~\rm{eV}~(\rm{CUORICINO}).
\label{25}
\ee
New experiments on the search for neutrinoless double $\beta$-decay
of different nuclei (CUORE ($^{130} \rm{Te}$), EXO ($^{136}
\rm{Xe}$), MAJORANA ($^{76} \rm{Ge}$), SuperNEMO ($^{82} \rm{Se}$),
MOON ($^{100} \rm{Mo}$) and others)  are in preparation \cite{bbfuture}.  In these
experiments large detectors (about one ton) will be used. In future
experiments on the search for $0\nu\beta\beta$-decay  the
sensitivity
\be |m_{ee}|\simeq {\rm{a~ few }\,~10^{-2}}~ \rm{eV}.\label{26}
\ee
are planned to be achieved.

\section{Effective Majorana mass and neutrino oscillation data}
The expected values of the effective Majorana neutrino mass,  which can be inferred from current 
neutrino oscillation data,  were considered in numerous papers (see \cite{predictions} and references therein). We will discuss here some issues.

For three neutrino masses two neutrino mass spectra are possible:

\begin{enumerate}
\item Normal spectrum
\be 
m_{1}<  m_{2}    <  m_{3} ;~ \Delta m^{2}_{12}  \ll    \Delta
m^{2}_{23}\label{27}
\ee
\item Inverted spectrum \footnote{In order to keep for solar-KamLAND
neutrino mass-squared difference notation $\Delta m^{2}_{12}$
different labeling of
neutrino masses are used   in the case of normal
and inverted spectra.
For
neutrino mixing angles in both cases the same notations can be used.}
\be 
m_{3}<  m_{1}    <  m_{2} ;~ \Delta m^{2}_{12}  \ll
 | \Delta m^{2}_{13}| \label{28}
\ee
\end{enumerate}
In the framework of the leading approximation it is not possible to distinguish normal and inverted spectra.
In order to reveal the type of the neutrino mass spectrum
necessary to study small effects beyond the leading approximation.
The size of such effects and possibilities to measure them depend on the value of the parameter
$\sin^{2}\theta_{13}$. We will demonstrate here that the effective Majorana mass
$m_{ee}$ strongly  depends on the type of the  neutrino
mass spectrum.

In the standard parametrization for the elements $U_{ei}$ 
we have
\be 
U_{e1}=\cos  \theta_{13}~\cos \theta_{12}~e^{i\beta_{1}};~     
U_{e2}=\cos  \theta_{13}~\sin \theta_{12}~e^{i\beta_{2}};~
U_{e3}=\sin \theta_{13} ~e^{i\beta_{3}}\label{29} 
\ee
The value of the angle $\theta_{12}$ is known  from
analysis of the data of the solar and KamLAND experiments (see
(\ref{12})). The angle $\theta_{13}$ is small and limited by the
CHOOZ bound (\ref{7})). Majorana phases $\beta_{i}$ are unknown.

From the analysis of the neutrino oscillation data two neutrino
mass-squared differences $\Delta m^{2}_{12}$ and $|\Delta
m^{2}_{23}|$ were determined. For neutrino masses in the case of the
normal spectrum we have
\be 
m_{2}=\sqrt{m^{2}_{1}+\Delta m^{2}_{12}},~~
m_{3}=\sqrt{m^{2}_{1}+\Delta m^{2}_{12}+\Delta
m^{2}_{23}}.\label{30}
\ee

In the case of the inverted spectrum neutrino masses are given by
\be 
m_{1}=\sqrt{m^{2}_{3}-\Delta m^{2}_{13}},~~
m_{2}=\sqrt{m^{2}_{3}-\Delta m^{2}_{13}+\Delta
m^{2}_{12}}.\label{31}
\ee
The lightest neutrino mass $m_{0}=m_{1}(m_{3})$ is
unknown at present. Upper bound of $m_{0}$, obtained from the data of the latest tritium
experiments, is given in (\ref{14}).

We will consider three characteristic neutrino mass spectra
\begin{enumerate}
\item Neutrino mass hierarchy
\be m_{1}\ll m_{2}\ll m_{3} \label{32}\ee
This type of the neutrino mass spectra is suggested by $SO(10)$ and
other GUT models which unify quarks and leptons (see \cite{Theory}). 

In the case of the hierarchy (\ref{32})  neutrino masses are
determined by the neutrino mass squared
differences and are known from the oscillation data:
\be 
m_{2}\simeq \sqrt{\Delta m^{2}_{\rm{12}}}\simeq 8.9\cdot
10^{-3}\rm{eV};~~ m_{3}\simeq \sqrt{\Delta m^{2}_{\rm{23}}}\simeq
4.9\cdot 10^{-2}\rm{eV}\label{33}
\ee
Neglecting the contribution of the lightest neutrino mass $m_{1}$
for the effective Majorana mass we obtain the following expression
\be
|m_{ee}|\simeq\left |\,
 \sin^{2} \theta_{12}\, \sqrt{\Delta m^{2}_{12}} + e^{i\,\beta_{32}}
 \sin^{2} \theta_{13}\, \sqrt{\Delta m^{2}_{23}}\,\right|,
 \label{34}
\ee
where $\beta_{32}=\beta_{3}- \beta_{2} $ is the Majorana phase 
difference.

The first term of (\ref{34}) is small because of the smallness of
$\sqrt{\Delta m^{2}_{12}}$. Contribution of ``large'' $\sqrt{\Delta
m^{2}_{32}}$ is suppressed by the small factor $\sin^{2} \theta_{13}
$. If we will use the  CHOOZ  bound (\ref{7}) the modulus of both terms
in (\ref{34}) are approximately equal. Hence the terms in (\ref{34})
could cancel each other and $|m_{ee}|$ could be  very small. From
(\ref{7}),  (\ref{9}) and (\ref{12}) for the upper bound of
$|m_{ee}|$ we find
\be 
|m_{ee}| \leq 6.4 \cdot 10^{-3}~\rm{eV}\label{35}
\ee
Thus, in the case of the hierarchy of neutrino masses, even upper
bound of the effective Majorana mass is about two times smaller that
the expected sensitivity of future experiments on the
search for $0\nu\beta\beta$-decay.
\item Inverted hierarchy of neutrino masses
\be 
m_{3}\ll m_{1} < m_{2}\label{36}
\ee
Such neutrino mass spectrum requires a
special flavor symmetry of the neutrino mass matrix. 
(for example, conservation of $L_{e}-L_{\mu}-L_{\tau}$).
For neutrino masses
in the case of the inverted hierarchy we have
\be 
m_{2}\simeq m_{1}\simeq \sqrt{ |\Delta m^{2}_{\rm{13}}|}; \,
m_{3}\ll \sqrt{ |\Delta m^{2}_{\rm{13}}|}\label{37}
\ee
Neglecting small contribution of the lightest neutrino mass, for the
effective Majorana mass we have the following expression
\be
|m_{ee}|\simeq \sqrt{| \Delta m^{2}_{\rm{13}}|}\,~ (1-\sin^{2}
2\,\theta_{\rm{12}}\,\sin^{2}\beta_{21})^{\frac{1}{2}}\label{38}
\ee
The only unknown parameter in (\ref{38}) is 
$\sin^{2}\alpha_{21}$.. This parameter  characterize CP violation in the case of
the Majorana neutrino mixing.

From the condition of the CP invariance in the lepton sector we have \cite{BNP}
\be 
U_{ei}=\eta_{i}\,U^{*}_{ei},\label{39}
\ee
where $\eta_{i}=\pm i$ is the CP parity of the Majorana neutrino
with the mass $m_{i}$. From (\ref{39}) for the Majorana CP phase
$\beta_{i}$ we find
\be 
2\,\beta_{i}=\frac{\pi}{2}~\rho_{i}+ 2 \pi n_{i}.\label{40}
\ee
Here $n_{i}$ is an integer number and $\rho_{i}=\pm 1$
is determined by the relation  $\eta_{i}=e^{i\frac{\pi}{2}\rho_{i}}$.
Thus, in the case of the CP invariance in the lepton sector from
(\ref{38}) and (\ref{40}) we find
\begin{itemize}
\item $|m_{ee}|_{CP1}=\sqrt{| \Delta m^{2}_{\rm{13}}|}$
(the same CP parities of $\nu_{2}$ and $\nu_{1}$)
\item $|m_{ee}|_{CP2}=\sqrt{| \Delta m^{2}_{\rm{13}}|}~\cos \theta_{12}$
(opposite CP parities of $\nu_{2}$ and $\nu_{1}$)

\end{itemize}

From (\ref{38}) for the effective Majorana mass we have the range
\be 
\cos  2\,\theta_{\rm{12}} \,\sqrt{ |\Delta m^{2}_{\rm{13}}|}
\leq |m_{ee}| \leq\sqrt{ |\Delta m^{2}_{\rm{13}}|}\label{41}
\ee
Upper and lower bounds in (\ref{41}) correspond to the case of the
CP conservation. Other values of the effective Majorana mass
corresponds to the case of the CP non conservation. 
Let us notice that the parameter
$\sin^{2}\beta_{21}$ is determined by the measurable quantities
\be 
\sin^{2}\beta_{21}=\frac{1}{\sin^{2}
2\,\theta_{\rm{12}}}~(1-\frac{|m_{ee}|^{2}}{| \Delta
m^{2}_{\rm{13}}|})\label{42}
\ee

From analysis of the solar oscillation data it was found that
$\theta_{\rm{12}}<\pi/4$ (see (\ref{12})). Thus, the lower bound of
the effective Majorana mass in (\ref{41}) is different from zero.
From (\ref{9}), (\ref{12})  and   (\ref{41})   we find the following 90 \% CL range
\be 
1.0\cdot 10^{-2}\leq |m_{ee}|\leq 5.5\cdot
10^{-2}~\rm{eV}\label{43}
\ee
The sensitivities to $|m_{ee}|$ of the most ambitious future
experiments on the search for $0\nu\beta\beta$ are in the range
(\ref{43}). Thus, next generation of the $0\nu\beta\beta$-
experiments could probe the inverted hierarchy of the neutrino
masses.

\item Quasi degenerate neutrino mass spectrum

If the lightest neutrino mass satisfies inequality
\be 
m_{\rm{1}}\gg \sqrt{ \Delta m^{2}_{\rm{23}}}~~~ (m_{\rm{3}}\gg
\sqrt{ |\Delta m^{2}_{\rm{13}}|})\label{44}
\ee
neutrino mass spectrum is practically degenerate
\be 
m_{1}\simeq m_{2}\simeq m_{3} \simeq m_{0}\label{45}
\ee
This spectrum requires symmetry of the neutrino mass matrix and only
marginally is compatible with neutrino oscillation data (see \cite{Altarelli}).

For the effective Majorana mass in the case of the quasi degenerate
spectrum we have
\be |m_{ee}|\simeq m_{0}\, (1-\sin^{2}
2\,\theta_{\rm{12}}\,\sin^{2}\alpha_{21})^{\frac{1}{2}}\label{46}\ee
There are two unknown parameters in Eq. (\ref{46}): $m_{0}$ and
$\sin^{2}\alpha_{21}$. From the measurement of $|m_{ee}|$ for the
lightest neutrino mass the following range can be obtained
\be
|m_{ee}|\leq m_{0}\leq  2.4\, |m_{ee}|\label{47}
\ee
The future tritium experiment KATRIN \cite{Katrin} will be sensitive to
$m_{0}\simeq 0.2$ eV.

\end{enumerate}

\section{Problem of nuclear matrix elements}

The observation of the $0\nu\beta\beta$-decay would be of  a profound
importance for our understanding of the origin of small neutrino
masses. The establishment of the Majorana nature of neutrinos with
definite masses would be a strong support of the most plausible
see-saw mechanism of neutrino mass generation, which connect the
smallness of neutrino masses with the violation of the total lepton
number at a large scale.

 If $0\nu\beta\beta$-decay would be observed, it will be very
important to obtain precise value of the effective Majorana mass
$|m_{ee}|$. As we have seen, the determination of $|m_{ee}|$
would allow to obtain an important information about neutrino mass
spectrum, mass of the lightest neutrino and, possibly, Majorana CP
phase difference.
From experimental data, however, only  {\em the product} of the
effective Majorana mass and nuclear matrix element  can be
determined. Nuclear matrix elements  must be calculated.

 The calculation of NME is a complicated nuclear problem (see reviews\cite{bbreviews2}). 
 NME
is  the matrix element of  an integral of the T-product of two hadronic charged weak currents and neutrino
propagator. Many intermediate nuclear states must be taken into
account in  calculations.

Two approaches, which are based on different
physical assumptions,  
are usually used for the calculation of NME:
Nuclear Shell Model (NSM) and Quasiparticle Random Phase
Approximation (QRPA). In literature exist many  QRPA-based  models. As
a result different calculations of the same NME differ by factor 2-3
or even more.

 We will discuss here  possible method which could allow to test NME
calculations in a model independent way \cite{BGrifols}.  We will use factorization
property of matrix elements of $0\nu\beta\beta$-decay which is based
on the assumption that Majorana neutrino mass mechanism is the
dominant mechanism of the $0\nu\beta\beta$-decay.

Several future experiments on the search for $0\nu\beta\beta$-decay of different nuclei will have comparable sensitivity to $|m_{ee}|$. Thus,
if $0\nu\beta\beta$-decay of one nuclei  will be discovered in a future experiment 
it is probable that the process will be observed also in other  experiments with different  nuclei.
The effective Majorana mass, which can be determined from the measurement of half-lives of the
$0\nu\beta\beta$-decay of different nuclei,  must be the same. From
this requirement we obtain the following
relations between half-lives of nuclei $A_{i},Z_{i}$ and
$A_{k},Z_{k}$
\be T_{1/2}(A_{i},Z_{i})=X_{M}(i;k)~T_{1/2}(A_{k},Z_{k}),
\label{48}
\ee
where coefficients $X_{M}(i;k)$ are given by the expression
\be
X_{M}(i;k)=\left(\frac{|M^{0\,\nu}(A_{k},Z_{k})|^{2}}
{|M^{0\,\nu}(A_{i},Z_{i})|^{2}}\right)_{M}~
\frac{G^{0\,\nu}(E^{k}_{0},Z_{k})}{G^{0\,\nu}(E^{i}_{0},Z_{i}}.\label{49}
\ee
The values of these coefficients depend on the model of the calculation of NME.
A model $M$ is compatible with data if relations (\ref{48}) are
satisfied. Let us stress, however, that this does not mean that the
model $M$ allows to obtain the correct value of the effective
Majorana mass.

For illustration we calculated coefficients $X_{M}(i;k)$ for three
latest models of NME calculations:
\begin{enumerate}
\item ($M_{1}$) Shell Model ( E. Courier et al.\cite{NSM} )

\item ($M_{2}$) QRPA (V. Rodin et al, \cite{Fedor};
important  QRPA parameter $g_{pp}$ 
is  fixed by the data of the experiments on the measurement of half-
lives of the $2\nu\beta\beta$-decay.)

\item ($M_{3}$) QRPA ( O. Civitarese, J. Suhonen  \cite{Civitarese}; parameters of the
QRPA model were fixed by the $\beta$-decay data of nearby nuclei)
\end{enumerate}
The results of the calculation are presented in the Table I.
\begin{center}
 Table I
\end{center}
\begin{center}
The coefficients $X_{M}(i;k)$, obtained in three recent models of
NME calculations ($M_{1}$ \cite{NSM}, $M_{2}$ \cite{Fedor}, $M_{3}\cite{Civitarese}$)
\end{center}

\begin{center}
\begin{tabular}{|c|ccc|}
\hline

&

$M_{1}$ & $M_{2}$ & $M_{3}$
\\
\hline & & & \\$X(^{100} \rm{Mo};^{76} \rm{Ge})$ &
& 0.59 & 0.17
\\ & & &  \\
$X(^{130} \rm{Te};^{76} \rm{Ge})$ & 0.25 & 0.49 & 0.13
\\ & & & \\
$X(^{136} \rm{Xe};^{76} \rm{Ge})$ & 0.55 & 0.80 & 0.07
\\ & & & \\
\hline
\end{tabular}
\end{center}

From the Table I we see that the measurement of
$0\nu\beta\beta$-decay of $^{76} \rm{Ge}$ and $^{130} \rm{Te}$ ( or  $^{76} \rm{Ge}$ and 
$^{136} \rm{Xe}$ or  $^{76} \rm{Ge}$ and
$^{100} \rm{Mo}$) 
can tell us which  model ($M_{1}$,  $M_{2}$ or $M_{3}$) is
compatible with data (if any). This conclusion, however, depends on
the choice of nuclei. Let us consider, for  example, the pair
$^{100} \rm{Mo}$ and $^{130} \rm{Te}$. We have
\be X(^{100} \rm{Mo};^{130} \rm{Te}) =1.2\,~(M_{2}); \,~~1.3\,~
(M_{3})\label{50}
\ee
Thus, if the relation (\ref{48})   for $^{100} \rm{Mo}$ and $^{130}
\rm{Te}$ is satisfied, say,  for the model $M_{2}$, it will be
difficult to exclude also the model $M_{3}$. However, the values of
the effective Majorana mass which can be obtained with the help of
these two models are quite different:
\be
|m_{ee}|_{M_{2}}= 2.6\,~ |m_{ee}|_{M_{3}}\label{51}
\ee
We come to the conclusion that the observation of
$0\nu\beta\beta$-decay of three (or more) nuclei would be an
important tool  for the test of the models of NME calculation and
for the determination of the value of the effective Majorana mass
$|m_{ee}|$.

\section{Conclusion}
The establishment of the nature of neutrinos with definite masses
$\nu_{i}$ (Majorana or Dirac?) will have a profound importance for
the understanding of the origin of small neutrino masses and
neutrino mixing. Investigation of the neutrinoless double
$\beta$-decay is the most sensitive probe of  the Majorana nature of
neutrinos. Today's limit on the effective Majorana mass is $|m_{ee}|
\leq (0.2-1.2)\,~\rm{eV}$.
The sensitivity $|m_{ee}| \simeq \rm{a\,~ few} ~10^{-2}\rm{eV}$ is
a challenging goal of future experiments.

 If $|m_{ee}|$ is determined the pattern of the neutrino mass
spectrum,  lightest neutrino mass and, possibly, Majorana CP phase
can be inferred. Calculation of nuclear matrix elements is a very
important, challenging nuclear problem. Observation of $0\nu
\beta\beta $-decay of several nuclei could allow to test  NME
calculations.

I acknowledge the support of  the Italien Program ``Rientro dei
cervelli''.

\end{document}